\documentclass[]{spie}  

\usepackage{amsmath,amsfonts,amssymb}
\usepackage{graphicx}
\usepackage{subcaption,dsfont}

\usepackage[colorlinks=true, allcolors=blue]{hyperref}

\title{A user model for JND-based video quality assessment: theory and applications}

\author[a]{Haiqiang Wang}
\author[b]{Ioannis Katsavounidis}
\author[a]{Xinfeng Zhang}
\author[a]{Chao Yang}
\author[a]{C.-C. Jay Kuo}
\affil[a]{University of Southern California, Los Angeles, California, USA}
\affil[b]{Netflix, Los Gatos, California, USA}

\authorinfo{Further author information: (Send correspondence to Haiqiang Wang)\\
Haiqiang Wang: E-mail: haiqianw@usc.edu}

\begin{document}
\maketitle

\begin{abstract}

The video quality assessment (VQA) technology has attracted a lot of
attention in recent years due to an increasing demand of video streaming
services. Existing VQA methods are designed to predict video quality in
terms of the mean opinion score (MOS) calibrated by humans in subjective
experiments. However, they cannot predict the satisfied user ratio
(SUR) of an aggregated viewer group. Furthermore, they provide little
guidance to video coding parameter selection, e.g. the Quantization
Parameter (QP) of a set of consecutive frames, in practical video
streaming services.  To overcome these shortcomings, the
just-noticeable-difference (JND) based VQA methodology has been proposed
as an alternative.  It is observed experimentally that the JND location
is a normally distributed random variable. In this work, we explain this
distribution by proposing a user model that takes both subject
variabilities and content variabilities into account. This model is
built upon user's capability to discern the quality difference between
video clips encoded with different QPs.  Moreover, it analyzes video
content characteristics to account for inter-content variability.  The
proposed user model is validated on the data collected in the VideoSet.
It is demonstrated that the model is flexible to predict SUR
distribution of a specific user group.

\end{abstract}


\section{Introduction}\label{sec:intro}

Although being expensive in time and money, the subjective experiment is
the ultimate method to quantify the perceptual quality of compressed
video.  Obtaining accurate and robust labels based on subjective votings
provided by human observers is a critical step in Quality of Experience
(QoE) evaluation. A typical subjective experiment
\cite{assembly2003methodology} involves: 1) selecting several
representative stimuli, 2) presenting them to a group of subjects and 3)
assigning quality scores to them by subjects. The collected subjective
scores should go through a cleaning and modeling process before being
used to validate the performance of objective video quality assessment
metrics.

Absolute Category Rating (ACR) is one of the most commonly used
subjective test methods. Test video clips are displayed on a screen for a
certain amount of time and observers rate their perceived quality using
an abstract scale \cite{itu1999subjective}, such as ``Excellent (5)'',
``Good (4)'', ``Fair (3)'', ``Poor (2)'' and ``Bad (1)''. There are two
approaches in aggregating multiple scores on a given clip. They are the
mean opinion score (MOS) and the difference mean opinion score (DMOS).
The MOS is computed as the average score from all subjects while the
DMOS is calculated from the difference between the raw quality scores of
the reference and the test images.

Both MOS and DMOS are popular in the quality assessment community.
However, they have several limitations \cite{chen2009crowdsourceable,
ye2014active}. The MOS scale is as an interval scale rather than an
ordinal scale.  It is assumed that there is a linear relationship
between the MOS distance and the cognitive distance. For example, a
quality drop from ``Excellent'' to ``Good'' is treated the same as that
from ``Poor'' to ``Bad''. There is no difference to a metric learning system
as the same ordinal distance is preserved ({\em i.e.} the quality distance is 1 for both cases in the aforementioned 5-level scale). However, human viewing
experience is quite different when the quality changes at different
levels.  It is also rare to find a video clip exhibiting poor or bad
quality in real-life video applications. As a consequence, the number of useful quality levels drops from five to three. It is too coarse for video quality measurement.

The second challenge is that scores from subjects are typically assumed
to be independently and identically distributed (i.i.d.) random
variables.  This assumption rarely holds. Given multiple quality votings
on the same content, individual voting contributes equally in the MOS
aggregation method \cite{whitehill2009whose}. Subjects may have
different levels of expertise on perceived video quality. A critical
viewer may give low quality ratings on coded clips whose quality is
still good to the majority \cite{li2014confidence}. The same phenomenon
occurs in all presented stimuli. The absolute category rating method is
confusing to subjects as they have different understanding and
interpretation of the rating scale.

To overcome the limitations of the MOS method, the
just-noticeable-difference (JND) based VQA methodology was proposed in
\cite{lin2015experimental} as an alternative. A viewer is asked to
compare a pair of coded clips and determine whether noticeable
difference can be observed or not. The pair consists of two stimuli,
i.e. a distorted stimulus (comparison) and an anchor preserving the
targeted quality. A bisection search is adopted to reduce the number of
pair comparisons.  The JND reflects the boundary of perceived quality
levels, which is well suited for the determination of the optimal
image/video quality with minimum bit rates. For example, the first JND,
whose anchor is the source clip, is the boundary between ``Excellent''
and ``Good'' categories. The boundary is subjectively decided rather
than empirically selected by the experiment designer.

In MOS or JND-based VQA methods, subjective data are noisy due to the
nature of ``subjective opinion''. In the extreme case, some subjects
submit random answers rather than good-faith attempts to label. Even
worse, adversary votings may happen due to malice or a systematic
misinterpretation of the task. Thus, it is critical to study subject
capability and reliability to alleviate their effects in the VQA task.

In this work, we propose a user model that takes subject bias and
inconsistency into account. The perceived quality of compressed video is
characterized by the satisfied user ratio (SUR). The SUR value is a
continuous random variable depending on subject and content factors. We
study the SUR difference as it varies with user profile as well as
content with variable level of difficulty. The proposed model
aggregates quality ratings per user group to address
inter-group difference.  The proposed user model is validated on the
data collected in the VideoSet\cite{wang2017videoset}.  It is demonstrated that the model is
flexible to predict SUR distribution of a specific user group.

The rest of this paper is organized as follows. Related work is reviewed
in Sec. \ref{sec:review}. The proposed user model is presented in Sec.
\ref{sec:user_model}.  Experimental results are shown in Sec.
\ref{sec:experiment}. Finally, concluding remarks are given in Sec.
\ref{sec:conclusion}.

\section{Related work}\label{sec:review}

There were several popular datasets available in the video quality
assessment community, such as LIVE \cite{seshadrinathan2010study},
VQEG-HD \cite{video2010report}, MCL-V \cite{Lin20151}, and NETFLIX-TEST
\cite{li2017recover}, using the MOS aggregation approach. Recently,
efforts have been made to examine MOS-based subjective test methods.
Various methods were proposed from different perspectives to address the
limitations mentioned in Section \ref{sec:intro}.

A theoretical subject model \cite{janowski2015accuracy} was proposed to
model the three major factors that influence MOS accuracy: subject bias,
subject inaccuracy, and stimulus scoring difficulty. It was reported
that the distribution of these three factors spanned about $\pm 25\%$ of
the rating scale. Especially, the subject error terms explained
previously observed inconsistencies both within a single subject's data
and also the lab-to-lab differences. A perceptually
weighted rank correlation indicator \cite{8272001} was proposed, which rewarded the
capability of corrected ranking high-quality images and suppressed the
attention towards insensitive rank mistakes. A
generative model \cite{li2017recover} was proposed to jointly recover content and subject
factors by solving a maximum likelihood estimation problem. However,
these models were proposed for the traditional MOS-based approaches.

Recently, there has been a large amount of efforts in JND-based video
quality analysis. The human visual system (HVS) cannot perceive small
pixel variation in coded video until the difference reaches a certain
level. However, the difference of selected contents for ranking in
traditional MOS-based framework was sufficiently large for the majority
of subjects. We could conduct fine-grained quality analysis by directly
measuring the JND threshold of each subject. There were several datasets
\cite{wang2017videoset,jin2016jndhvei,mcl_jcv} proposed with the JND
methodology. Corresponding JND prediction methods were proposed in
\cite{huang2017measure,wang2017prediction}. However, the JND location
was analyzed in a data-driven fashion. It was simply modeled by the mean
value of multiple JND samples with heuristic subject rejection approach.

A probability model \cite{wang2018jnd} was proposed to offer new
insights to the JND phenomenon. Inspired by \cite{li2017recover}, the
proposed generative model decomposed JND-based video quality score into
subject and content factors. A close-form expression was derived to
estimate the JND location by aggregating multiple binary decisions. It
was shown that the JND samples followed Normal distribution which was
parameterized by the subject and content factors. These unknown factors
were jointly optimized by solving a maximum likelihood estimation (MLE)
problem.

\section{Proposed user model}\label{sec:user_model}

In this section, we present the proposed user model based on the JND
methodology. Let $c$ denote a reference video content, which can be
compressed into a set of clips $e_{i}$, $i=0, 1, 2, \cdots, 51$, where
$i$ is the quantization parameter (QP) index used in H.264/AVC.
Typically, clip $e_{i}$ has a higher PSNR value than clip $e_{j}$, if
$i<j$, and $e_{0}$ is the losslessly coded copy of $c$.

The JND of coded clips characterizes the distortion visibility threshold
with respect to a given anchor, $e_{i}$.  Through the subjective
experiment, JND points can be obtained from a sequence of consecutive
Noticeable/Unnoticeable difference tests between clips pair $(e_{i},
e_{j})$ where $ j \in \{i+1, \cdots, 51\}$. For example, the anchor for
the first JND point is $e_{0}$ and it remains the same while searching
for the first JND point. A bisection search is adopted to effectively
update $e_{j}$ and reduce the total number of comparisons.

Consider a VQA dataset consisting of $C$ contents and $S$ subjects, the
JND data matrix is modeled as $Y \in \mathbb{R}^{C \times S}$.
Individual JND location $Y_{c,s}$ for $s=1, \cdots, S$ and $c=1, \cdots,
C$, is obtained through six rounds of comparison. The following analysis
is conducted on the data matrix to recover underlying subject and
content factors.

It was demonstrated in \cite{wang2018jnd} that the perceived video quality
depends on several causal factors: 1) the bias of the subject bias, 2)
the inconsistency of a subject, 3) the average JND location, 4) the
difficulty of a content to evaluate. The JND location of content $c$
from subject $s$ can be expressed as
\begin{equation} \label{eq:final_decom}
Y_{c, s} = y_{c} + \mathcal{N}(0, v_{c}^{2}) + \mathcal{N}(b_{s}, v_{s}^{2}),
\end{equation}
where $y_{c}$ and $v_{c}^{2}$ are content factors while $b_{s}$ and
$v_{s}^{2}$ are subject factors. The difficulty of a content is modeled
by $v_{c}^{2} \in [0, \infty)$. A larger $v_{c}^{2}$ value means that
its masking effects are stronger and the most experienced experts still
have difficulty in spotting artifacts in compressed clips. The bias of a
subject is modeled by parameter $b_{s} \in(-\infty, +\infty)$. If
$b_{s}<0$, the subject is more sensitive to quality degradation in
compressed video clips. If $b_{s}>0$, the subject is less sensitive to
distortions. The sensitivity of an averaged subject has a bias around
$b_{s}=0$. Moreover, the subject variance, $v_{s}^{2}$, captures the
inconsistency of the quality votings from subject $s$. A consistent
subject evaluate all sequences attentively.

\subsection{Satisfied user ratio on a specific user group}

Under the assumption that content and subject are independent factors on
perceived video quality, the JND position can be expressed by a Gaussian
distribution in form of
\begin{equation}\label{eq:jnd_gaussian}
Y_{c, s} \sim \mathcal{N} (\mu_{Y}, \sigma_{Y}^{2}),
\end{equation}
where $\mu_{Y}=y_{c}+b_{s}$ and $\sigma_{Y}^{2}=v_{c}^{2}+v_{s}^{2}$.
The unknown parameters are $\theta=(\{y_c\}, \{v_{c}\}, \{b_s\},
\{v_{s}\})$ for $c = 1, \cdots, C$ and $s = 1, \cdots, S$, where
$\{\cdot\}$ denotes the corresponding parameter set. All unknown
parameters can be jointly estimated via the Maximum Likelihood
Estimation (MLE) method given the subjective data matrix $Y \in
\mathbb{R}^{C \times S}$.  This is a well-formulated parameter inference
approach and we refer interested viewers to \cite{li2017recover,
wang2018jnd} for more details.

Among the four parameters $\theta=(\{y_c\}, \{v_{c}\}, \{b_s\},
\{v_{s}\})$, we have limited control on content factors, i.e. $y_{c}$
and $v_{c}$. Content factors should be independent parameters that are
input to a quality model. In practice, it is difficult, sometimes even
impossible, to model subject inconsistency ({\em i.e.}, the $v_{s}$
term), as it is viewer's freedom to decide how much attention to pay to
the video content.

On the other hand, the subject bias term ({\em i.e.} $b_{s}$) is a consistent
prior of each subject. It is reasonable to model the subject bias and
integrate it into a SUR model. We can roughly classify users into
three groups based on the bias estimated from MLE. The user model aims
to provide a flexible system to accommodate different viewer groups:
\begin{itemize}
\item Viewers who are easy-to-satisfy (ES), corresponding to a larger $b_{s}$;
\item Viewers who have normal sensitivity (NS), corresponding to a neural $b_{s}$;
\item Viewers who are hard-to-satisfy (HS), corresponding to a smaller $b_{s}$.
\end{itemize}

\begin{figure*}[!t]
\centering
  \begin{subfigure}[b]{0.19\linewidth}\includegraphics[width=1.0\linewidth]{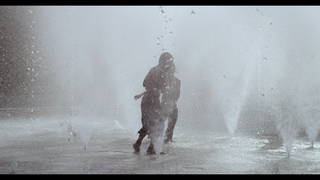}\end{subfigure}
  \begin{subfigure}[b]{0.19\linewidth}\includegraphics[width=1.0\linewidth]{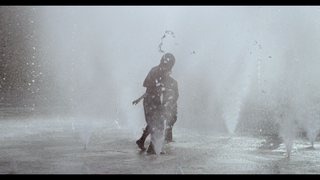}\end{subfigure}
  \begin{subfigure}[b]{0.19\linewidth}\includegraphics[width=1.0\linewidth]{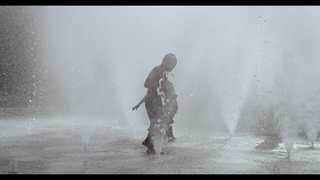}\end{subfigure}
  \begin{subfigure}[b]{0.19\linewidth}\includegraphics[width=1.0\linewidth]{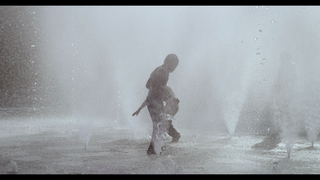}\end{subfigure}
  \begin{subfigure}[b]{0.19\linewidth}\includegraphics[width=1.0\linewidth]{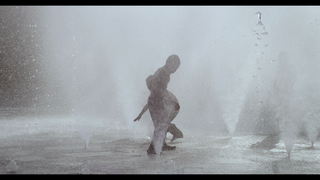}\end{subfigure}\\
  \begin{subfigure}[b]{0.19\linewidth}\includegraphics[width=1.0\linewidth]{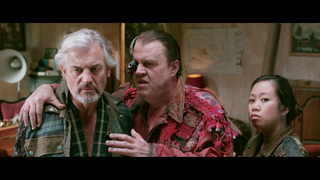}\end{subfigure}
  \begin{subfigure}[b]{0.19\linewidth}\includegraphics[width=1.0\linewidth]{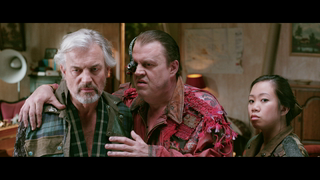}\end{subfigure}
  \begin{subfigure}[b]{0.19\linewidth}\includegraphics[width=1.0\linewidth]{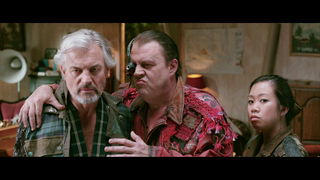}\end{subfigure}
  \begin{subfigure}[b]{0.19\linewidth}\includegraphics[width=1.0\linewidth]{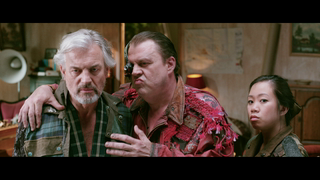}\end{subfigure}
  \begin{subfigure}[b]{0.19\linewidth}\includegraphics[width=1.0\linewidth]{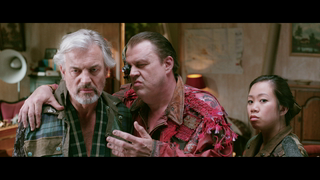}\end{subfigure}\\
\caption{Consecutive frames of contents $\#11$ (top) and $\#203$ (bottom), respectively.} \label{fig:thumbnails}
\end{figure*}

Furthermore, a viewer is said to be satisfied if one cannot perceive
quality difference between the compressed clip and its anchor. The
Satisfied User Ratio (SUR) of video clip $e_{i}$ on user group $j$ can
be expressed as
\begin{equation}\label{eq:sur_sum}
Z_{i,j} = 1-\frac{1}{|S_{j}|}\sum_{s\in S_{j}}\mathds{1}_{s}(e_{i}),
\end{equation}
where $S_{j}$ is the $j-$th group of subjects and $|\cdot|$ denotes the
cardinality. $\mathds{1}_{s}(e_{i})=1$ or $0$ if the $s-$th subject can
or cannot see the difference between compressed clip $e_{i}$ and its
anchor, respectively. The summation term in the right-hand-side of Eq.
(\ref{eq:sur_sum}) is the empirical cumulative distribution function
(CDF) of random variable $Y_{c,s}$. Then, by substituting Eq.
(\ref{eq:jnd_gaussian}) into Eq. (\ref{eq:sur_sum}), we obtain a
compact expression for the SUR curve as
\begin{equation}\label{eq:sur_q}
Z_{i,j} = Q(e_{i}|\mu_{Y}, \sigma_{Y}^{2}) = Q(e_{i}|y_{c}+b_{s},
v_{c}^{2}+v_{s}^{2}), \quad \text{for} \quad s \in S_{j},
\end{equation}
where $Q(\cdot)$ is the Q-function of the normal distribution. By
dividing users into different groups, the model achieves small
intra-group variance and large inter-group variance. We can model JND
and SUR more precisely. Alternatively, a universal model could be
generalized by replacing $S_{j}$ by the union of all subjects, {\em
i.e.} $S = \bigcup_{j} S_{j}$.

\section{Experimental results}\label{sec:experiment}

We evaluate the performance of the proposed user model using real JND
data from the VideoSet \cite{wang2017videoset} and compare it with the
MOS method. The VideoSet contains 220 video contents in four resolutions
and three JND points per resolution per content. During the subjective
test, the dataset was split into 15 subsets and each subset was
evaluated independently by a group of subjects. The group size was
around 35. We adopt a subset of the first JND point on 720p video in our
experiment.  It contains 15 video contents evaluated by 37 subjects.

\begin{figure*}[!t]
\centering
	\begin{subfigure}[b]{0.45\linewidth}
	\centering{}
	\includegraphics[width=1.0\linewidth]{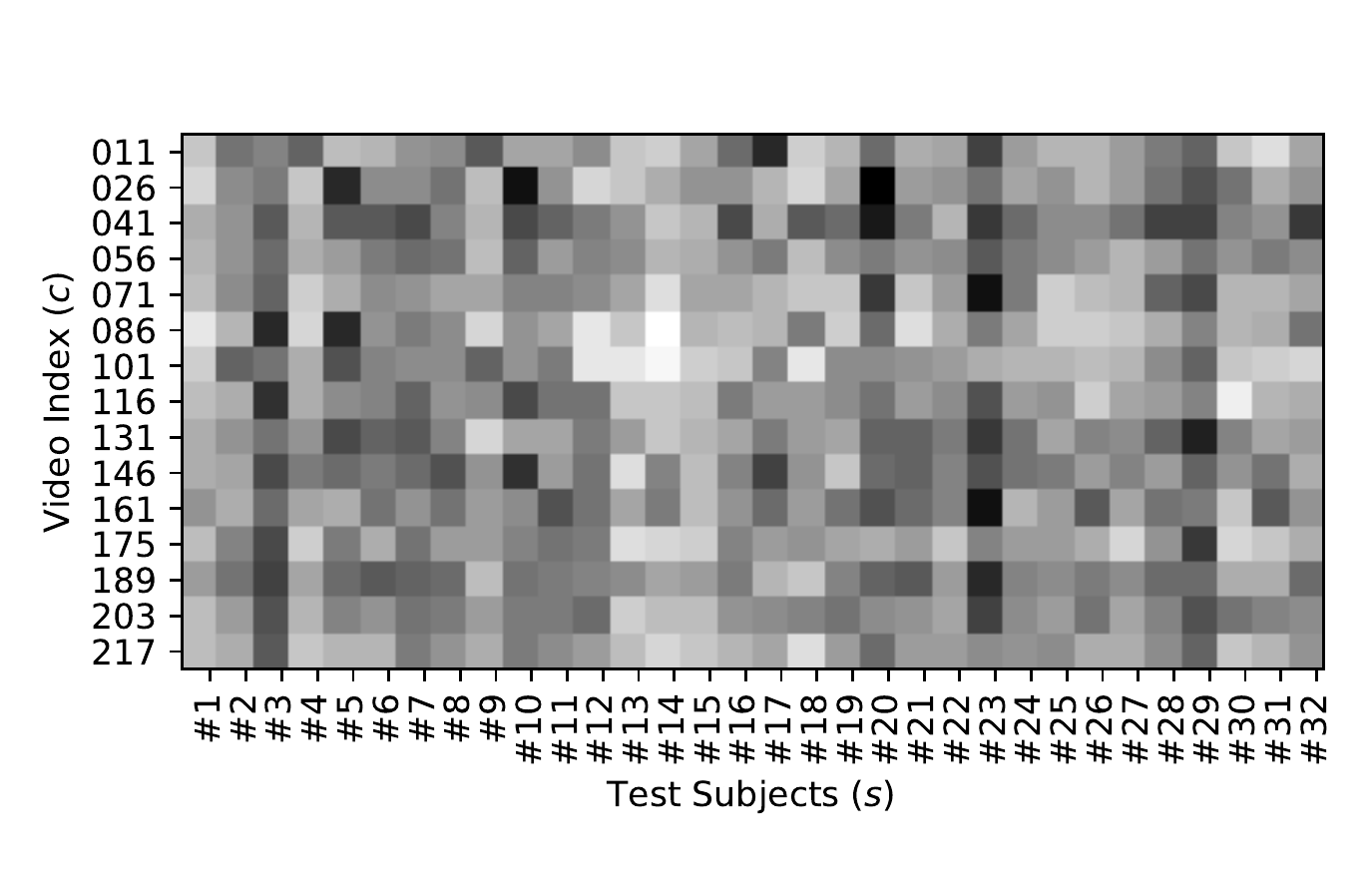}
  \caption{\label{fig:clean_ycs}}
	\end{subfigure}
  \begin{subfigure}[b]{0.45\linewidth}
	\centering{}
  \includegraphics[width=1.0\linewidth]{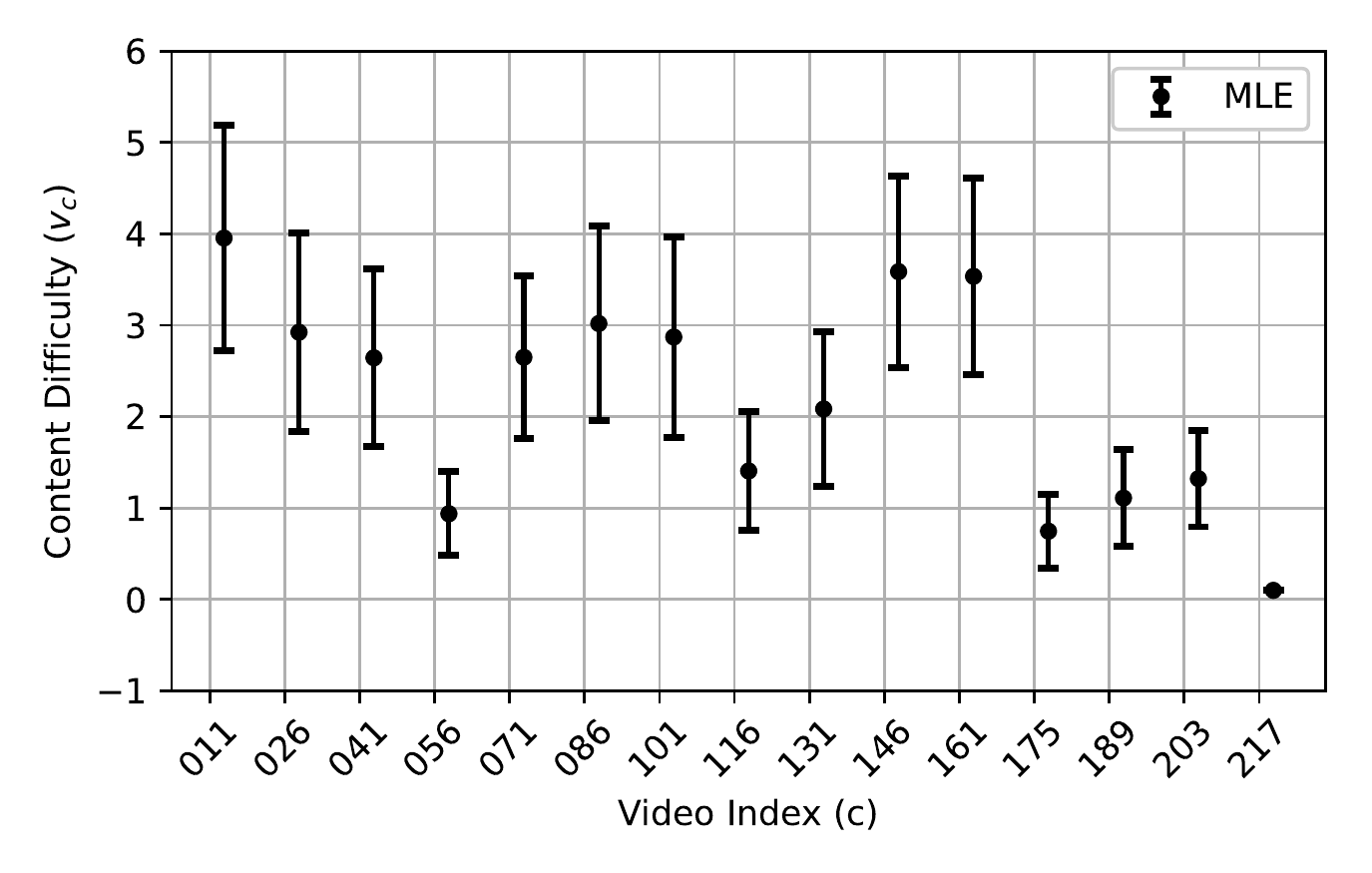}
  \caption{\label{fig:clean_content_difficulty}}
	\end{subfigure}
  \begin{subfigure}[b]{0.45\linewidth}
	\centering{}
	\includegraphics[width=1.0\linewidth]{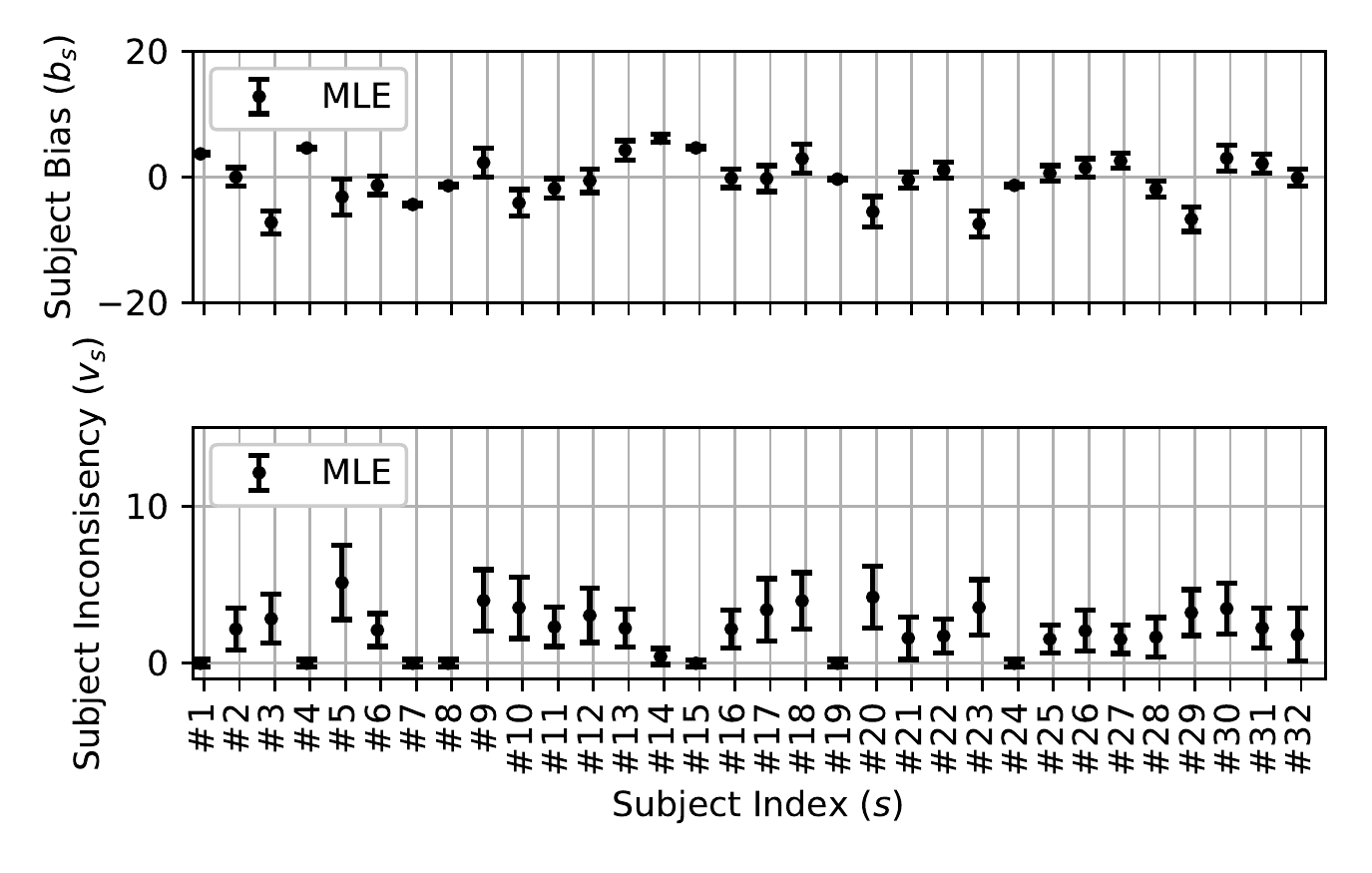}
  \caption{\label{fig:clean_subj}}
	\end{subfigure}
  \begin{subfigure}[b]{0.45\linewidth}
	\centering{}
	\includegraphics[width=1.0\linewidth]{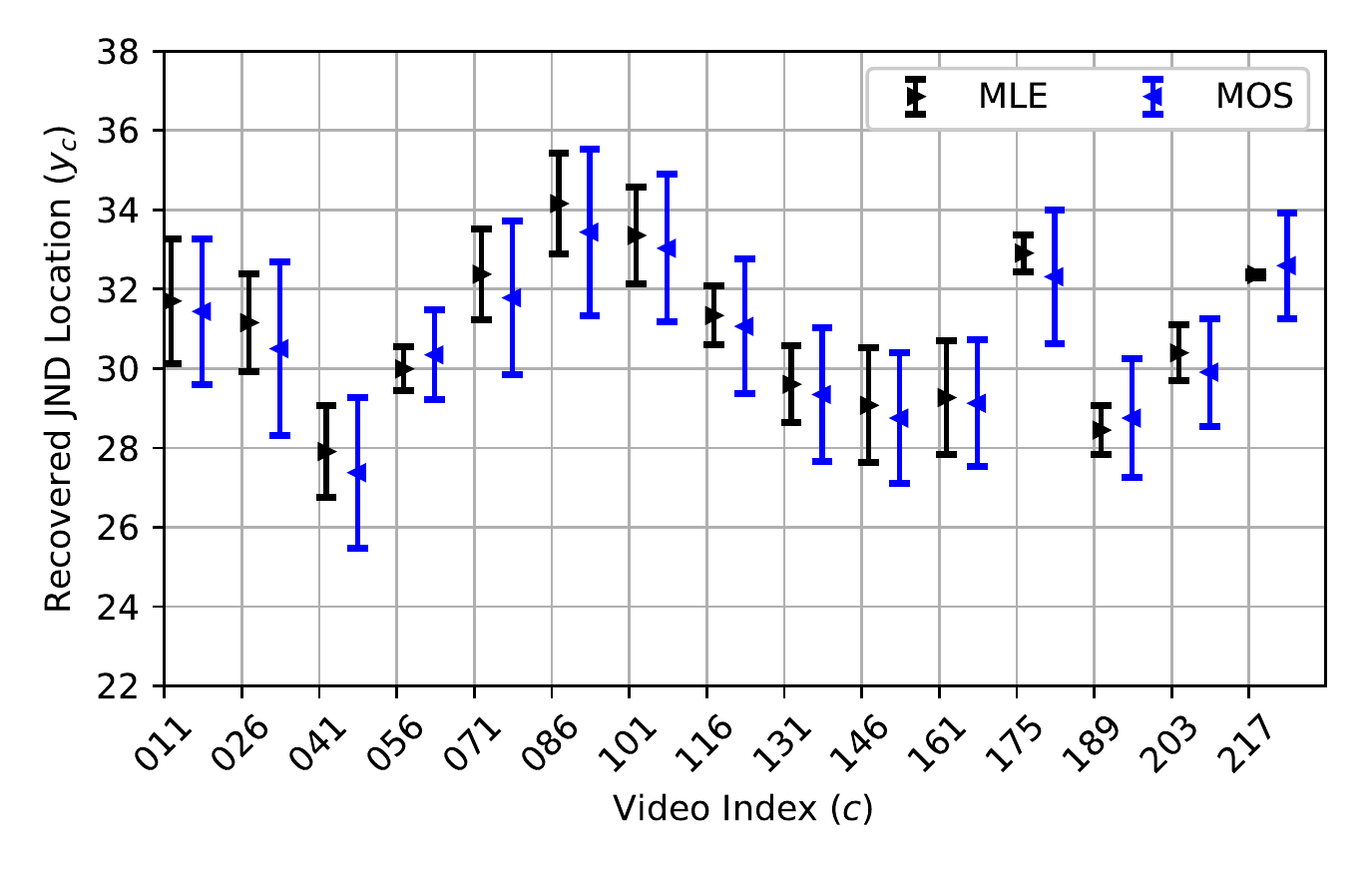}
  \caption{\label{fig:clean_yc_mle_mos}}
	\end{subfigure}
\caption{Visualization of cleaned JND data and estimated subject and
content factors: (a) cleaned JND data, where each pixel represents one
JND location and a brighter pixel means the JND happens at a larger QP,
(b) estimated content difficulty ({\em i.e.} $v_{c}$) using the MLE method, (c) estimated
subject bias and inconsistency ({\em i.e.} $v_{s}$), and (d) estimated JND locations using
the MLE and the MOS methods, respectively. The error bars in subfigures
represent $95\%$ confidence interval.} \label{fig:clean_subject_factor}
\end{figure*}

\subsection{Parameter Inference}

The cleaned JND scores are shown in Figure \ref{fig:clean_ycs} and the
estimated subject bias and inconsistency are shown in Figure
\ref{fig:clean_subj}, respectively. Please note that 5 subjects were
identified as unreliable subjects and their quality votings were
removed. These subjects have a larger bias value or inconsistent
measures. We refer interested readers to \cite{wang2018jnd} for further
details.

Figure \ref{fig:clean_content_difficulty} shows the estimated content
difficulty. Content $\#11$ is a scene about toddlers playing in a
fountain. The masking effect is strong due to water drops in the
background and moving objects. Thus, compression artifacts are difficult
to perceive, and it has the highest content difficulty. On the other
hand, content $\#203$ is a scene captured by a still camera. It focuses
on speakers with blurred still background. The content difficulty is low
as the masking effect is weak, and compression artifacts are more
noticeable. Representative frame thumbnails are given in Fig.
\ref{fig:thumbnails}.

The estimated JND location using the MLE method and the MOS method are
compared in Figure \ref{fig:clean_yc_mle_mos}. The MLE approach offers
more reliable estimation as its confidence intervals are much tighter
than those estimated by the MOS method.

\subsection{SUR on different viewer groups}\label{subsec:viewer_groups}

\begin{figure}[!t]
\centering
\begin{subfigure}[b]{1.0\linewidth}
  \centering
  \includegraphics[width=0.32\linewidth]{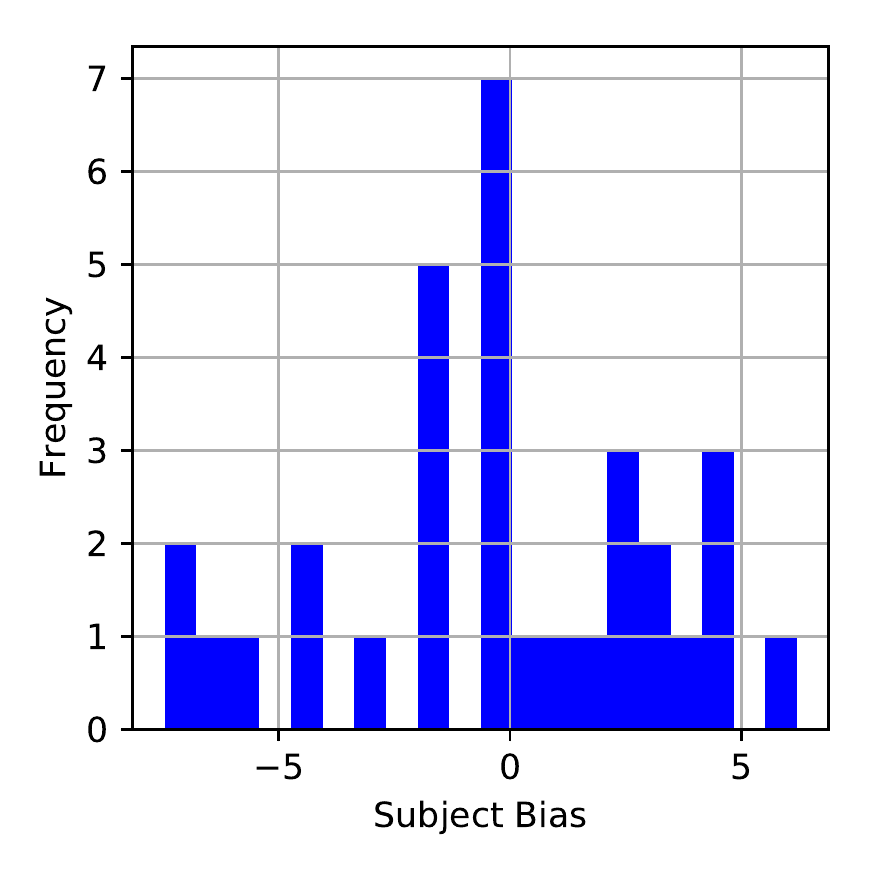}
  \includegraphics[width=0.32\linewidth]{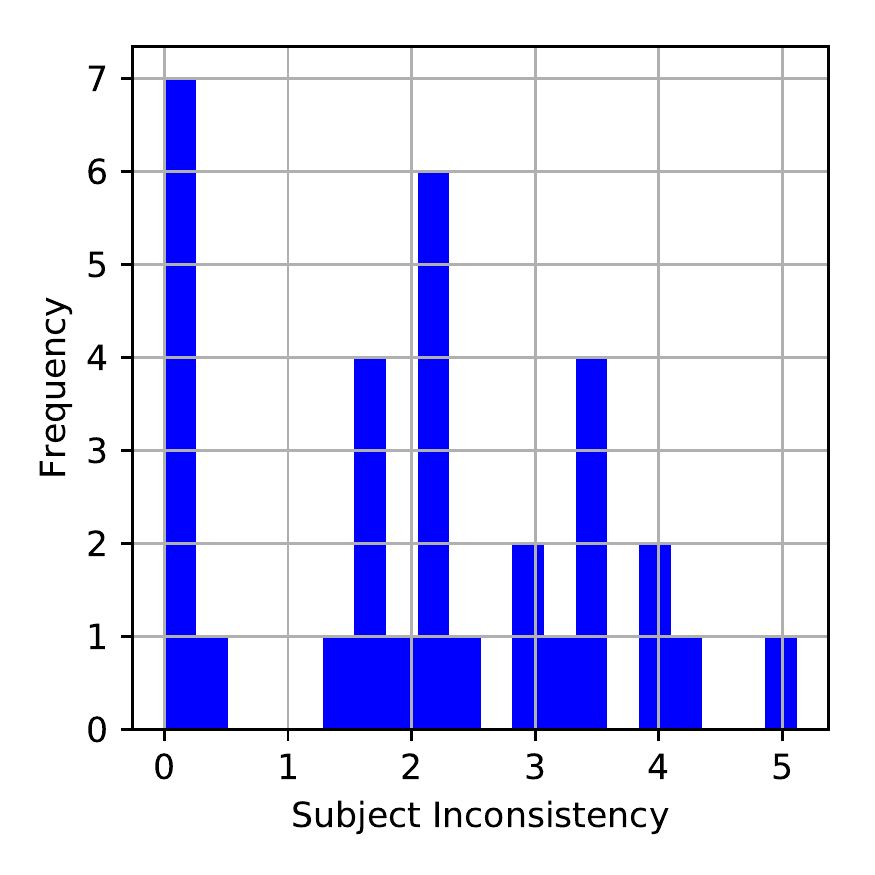}
  \includegraphics[width=0.32\linewidth]{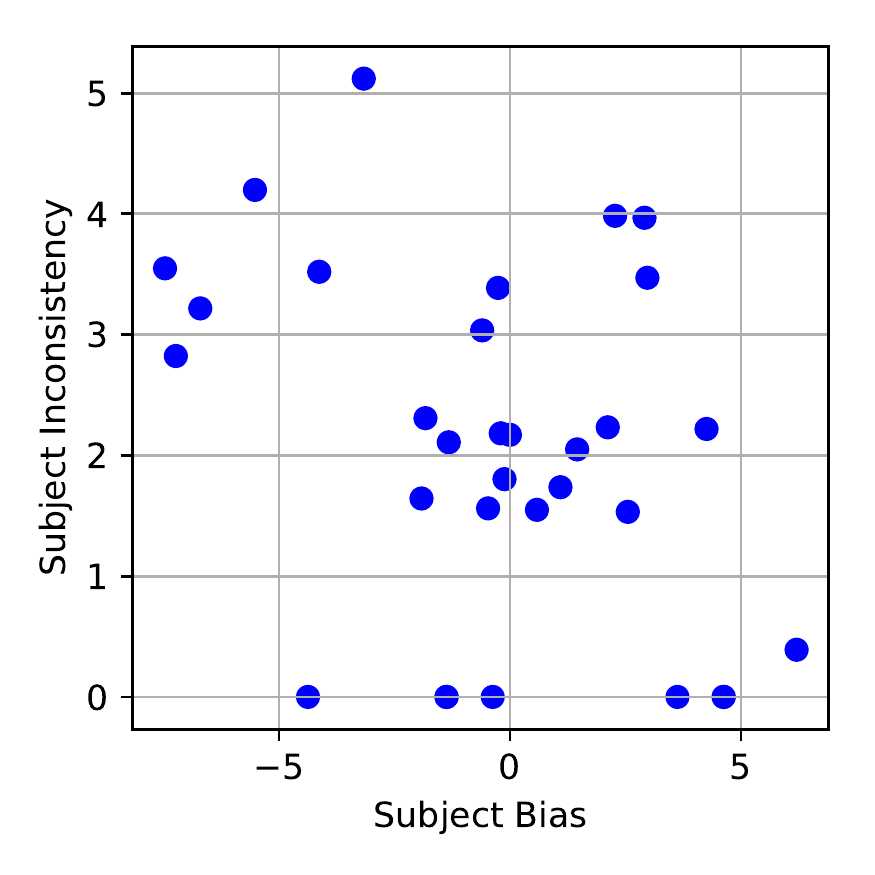}
\end{subfigure}
\caption{Illustration of subject factors. Left: the histogram of the
subject bias. Middle: the histogram of subject inconsistency. Right:
the scatter plot of subject inconsistency versus the subject bias.}
\label{fig:subj_factor}
\end{figure}

We classify viewers into different viewer groups based on the estimated
subject bias from cleaned JND data. The distribution of subject bias and
inconsistency are given in Figure \ref{fig:subj_factor}. The left and
middle figures are the histogram of their statistics, respectively. For a
large percentage of viewers, their bias and inconsistency are in a
reasonable range ({\em i.e.} $[-4, 4]$ for the subject bias and $[0, 3.5]$ for
subject inconsistency, respectively). The right figure is the scatter
plot of these two factors. We do not observe strong correlation between
them.

In the following, we use video $\#11$ and $\#203$ as input contents to
demonstrate the effectiveness of the proposed user model. Under fixed
content factors, we compare the SUR differences between different viewer
groups. Content $\#11$ has a strong masking effect so that it is
difficult to evaluate (HC, ``Hard Content'').  Content $\#203$ has a
weak masking effect and it is easy to evaluate (EC, ``Easy Content'').

\begin{figure}[!t]
\centering
\begin{subfigure}[b]{0.7\linewidth}
  \centering
  \includegraphics[width=0.9\linewidth]{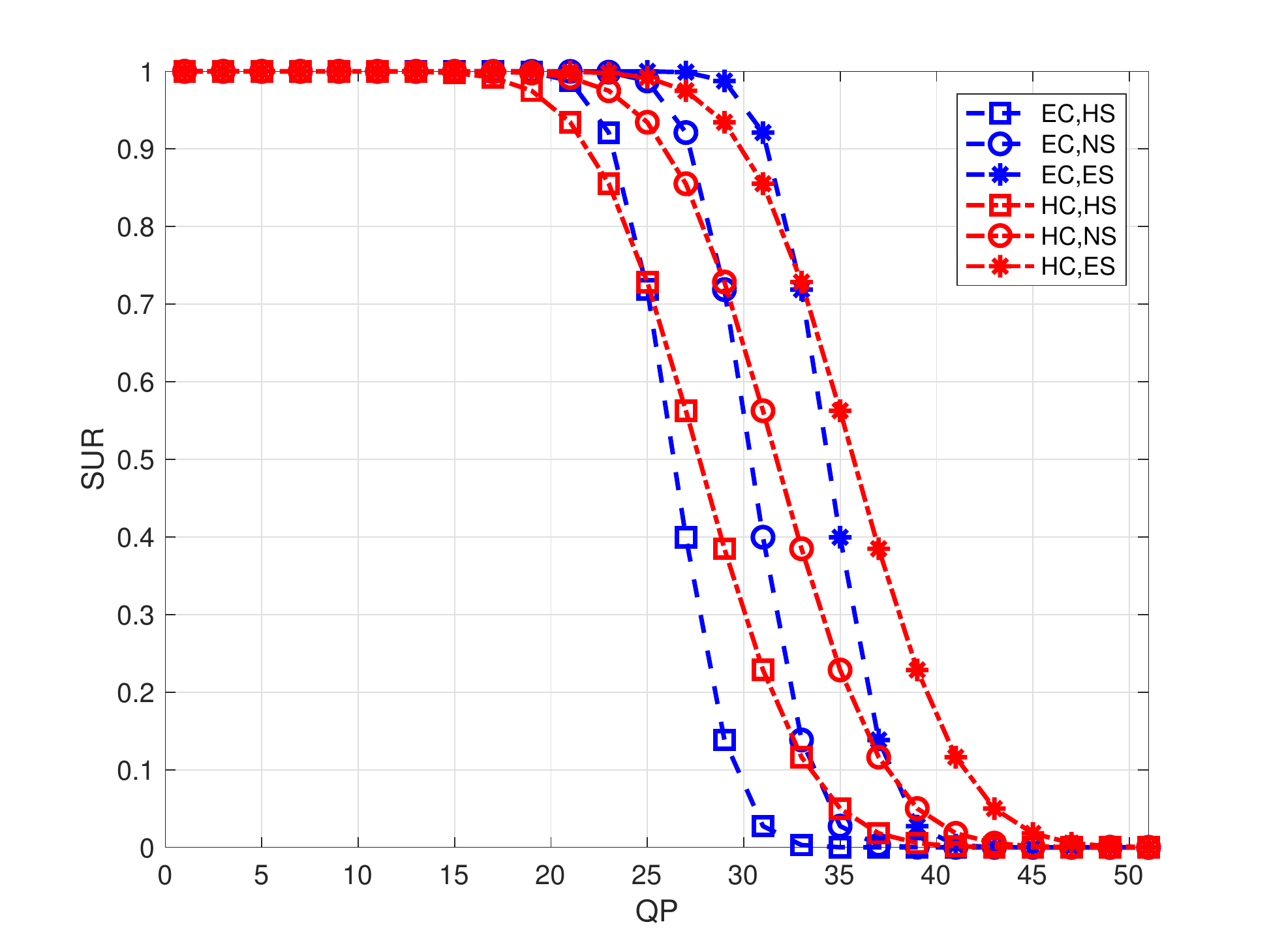}
\end{subfigure}
\caption{Illustration of the proposed user model. The blue and red
curves demonstrate the SUR of EC and HC contents, respectively. For each
content, the three curves show the SUR difference between different user
groups.} \label{fig:rst_1}
\end{figure}

Figure \ref{fig:rst_1} shows the effect of subject factors on the SUR
curve. There are 6 SUR curves by combining different content factors and
subject factors. The input parameters are obtained from MLE. They are
set as follows.
\begin{itemize}
\item The subject bias is set to -4, 0, 4 for HS, NS and ES, respectively.
\item Subject inconsistency is set to 2 for all subjects.
\item The averaged JND locations are set to 31.7 and 30.39 for clip $\#11$ and
$\#203$, respectively.
\item The content difficulty levels are set to 3.962 and 1.326 for clip $\#11$ and
$\#203$, respectively.
\end{itemize}

We have the following two observations.
\begin{enumerate}
\item SUR difference for normal users \\
Consider the middle curves of EC and HC contents. Subjects in this group
have normal sensitivity and we use this group to represent the majority.
Intuitively, the content diversity is large if we visually examine those
two clips. However, if we target at $SUR=0.75$, which is the counterpart of the mean value in the
MOS method, the QP location from modeled SUR curve is pretty close. The
difference increases when the SUR deviates from the $SUR=0.75$ location.
For contents that have a weak masking effect (shown in blue curve), they
are less resistant to compression distortion and SUR drops sharply once
artifacts become noticeable. In contrast, for contents that have a
strong masking effect (shown in red curve), they have better
discriminatory power on subject capability so that the SUR curve drops
slowly. Given the same extra bitrate quota, we could expect a higher SUR
gain from EC than HC. It takes much more effort to satisfy critical
users when the content has a strong masking effect. We conclude that it
is essential to study content difficulty and subject capability to
better model perceived quality of compressed video.
\item SUR difference for different user groups \\
The SUR difference is considerably large among different user groups on
the same content. We observe a gap between the three curves for both
contents. The SUR curve of normal users is shifted by the subject bias
$b_{s}$ in Eq. (\ref{eq:sur_q}). Although the neutral user group covers
the majority of users, we believe that a quality model would better
characterize QoE by taking the user capability into consideration.
\end{enumerate}

The above observations can be easily explained using the proposed user
model. It shows the value and power of our study.

\section{Conclusion and future work}\label{sec:conclusion}

A flexible user model was proposed in this work by considering the
subject and content factors in the JND framework. The QoE of a group of
users was characterized by the Satisfied User Ratio (SUR) while the JND
location of content $c$ from subject $s$ was modeled as a random
variable parameterized by subject and content factors. The model
parameters can be estimated by the MLE method using a set of JND-based
subjective test data. As an application of the proposed user model, we
studied SUR curves that are influenced by different user profiles
and contents of different difficult levels. It was shown that the subject
capability significantly affects the SUR curves, especially at the
middle range of the quality curve.

Apparently, the proposed user model provides valuable insights on the
quality assessment problem.  We would like to explore these insights for
better SUR prediction for new contents in the future.

\bibliography{report} 

\begin{thebibliography}{10}

\bibitem{assembly2003methodology}
{ITU-R BT. 500}, ``Methodology for the subjective assessment of the quality of
  television pictures,'' (2003).

\bibitem{itu1999subjective}
{ITU-T P.910}, ``Subjective video quality assessment methods for multimedia
  applications,'' (1999).

\bibitem{chen2009crowdsourceable}
Chen, K.-T., Wu, C.-C., Chang, Y.-C., and Lei, C.-L., ``A crowdsourceable {QoE}
  evaluation framework for multimedia content,'' in [{\em Proceedings of the
  17th ACM international conference on
  Multimedia}{\nolinebreak\hspace{0.1em}]},   491--500, ACM (2009).

\bibitem{ye2014active}
Ye, P. and Doermann, D., ``Active sampling for subjective image quality
  assessment,'' in [{\em Proceedings of the IEEE Conference on Computer Vision
  and Pattern Recognition}{\nolinebreak\hspace{0.1em}]},   4249--4256 (2014).

\bibitem{whitehill2009whose}
Whitehill, J., Wu, T.-f., Bergsma, J., Movellan, J.~R., and Ruvolo, P.~L.,
  ``Whose vote should count more: Optimal integration of labels from labelers
  of unknown expertise,'' in [{\em Advances in neural information processing
  systems}{\nolinebreak\hspace{0.1em}]},   2035--2043 (2009).

\bibitem{li2014confidence}
Li, Q., Li, Y., Gao, J., Su, L., Zhao, B., Demirbas, M., Fan, W., and Han, J.,
  ``A confidence-aware approach for truth discovery on long-tail data,'' {\em
  Proceedings of the VLDB Endowment}~{\bf 8}(4),  425--436 (2014).

\bibitem{lin2015experimental}
Lin, J.~Y., Jin, L., Hu, S., Katsavounidis, I., Li, Z., Aaron, A., and Kuo,
  C.-C.~J., ``Experimental design and analysis of {JND} test on coded
  image/video,'' in [{\em SPIE Optical Engineering+
  Applications}{\nolinebreak\hspace{0.1em}]},   95990Z--95990Z, International
  Society for Optics and Photonics (2015).

\bibitem{wang2017videoset}
Wang, H., Katsavounidis, I., Zhou, J., Park, J., Lei, S., Zhou, X., Pun, M.-O.,
  Jin, X., Wang, R., Wang, X., et~al., ``Videoset: A large-scale compressed
  video quality dataset based on {JND} measurement,'' {\em Journal of Visual
  Communication and Image Representation}~{\bf 46},  292--302 (2017).

\bibitem{seshadrinathan2010study}
Seshadrinathan, K., Soundararajan, R., Bovik, A.~C., and Cormack, L.~K.,
  ``Study of subjective and objective quality assessment of video,'' {\em IEEE
  Transactions on Image Processing}~{\bf 19}(6),  1427--1441 (2010).

\bibitem{video2010report}
Group, V. Q.~E. et~al., ``Report on the validation of video quality models for
  high definition video content,'' {\em VQEG, Geneva, Switzerland, Tech.
  Rep.[Online]. Available: http://www. its. bldrdoc.
  gov/vqeg/projects/hdtv/hdtv. aspx}  (2010).

\bibitem{Lin20151}
Lin, J.~Y., Song, R., Wu, C.-H., Liu, T., Wang, H., and Kuo, C.-C.~J.,
  ``{MCL-V}: A streaming video quality assessment database,'' {\em Journal of
  Visual Communication and Image Representation}~{\bf 30},  1 -- 9 (2015).

\bibitem{li2017recover}
Li, Z. and Bampis, C.~G., ``Recover subjective quality scores from noisy
  measurements,'' in [{\em Data Compression Conference (DCC),
  2017}{\nolinebreak\hspace{0.1em}]},   52--61, IEEE (2017).

\bibitem{janowski2015accuracy}
Janowski, L. and Pinson, M., ``The accuracy of subjects in a quality
  experiment: A theoretical subject model,'' {\em IEEE Transactions on
  Multimedia}~{\bf 17}(12),  2210--2224 (2015).

\bibitem{8272001}
Wu, Q., Li, H., Meng, F., and Ngan, K.~N., ``A perceptually weighted rank
  correlation indicator for objective image quality assessment,'' {\em IEEE
  Transactions on Image Processing}~{\bf 27},  2499--2513 (May 2018).

\bibitem{jin2016jndhvei}
Jin, L., Lin, J.~Y., Hu, S., Wang, H., Wang, P., Katasvounidis, I., Aaron, A.,
  and Kuo, C.-C.~J., ``Statistical study on perceived {JPEG} image quality via
  {MCL-JCI} dataset construction and analysis,'' in [{\em IS\&T/SPIE Electronic
  Imaging}{\nolinebreak\hspace{0.1em}]},  International Society for Optics and
  Photonics (2016).

\bibitem{mcl_jcv}
Wang, H., Gan, W., Hu, S., Lin, J.~Y., Jin, L., Song, L., Wang, P.,
  Katsavounidis, I., Aaron, A., and Kuo, C.-C.~J., ``{MCL-JCV}: A {JND}-based
  {H.264/AVC} video quality assessment dataset,'' in [{\em 2016 IEEE
  International Conference on Image Processing
  (ICIP)}{\nolinebreak\hspace{0.1em}]},   1509--1513 (Sept 2016).

\bibitem{huang2017measure}
Huang, Q., Wang, H., Lim, S.~C., Kim, H.~Y., Jeong, S.~Y., and Kuo, C.-C.~J.,
  ``Measure and prediction of hevc perceptually lossy/lossless boundary {QP}
  values,'' in [{\em Data Compression Conference (DCC),
  2017}{\nolinebreak\hspace{0.1em}]},   42--51, IEEE (2017).

\bibitem{wang2017prediction}
Wang, H., Katsavounidis, I., Huang, Q., Zhou, X., and Kuo, C.-C.~J.,
  ``Prediction of satisfied user ratio for compressed video,'' {\em arXiv
  preprint arXiv:1710.11090}  (2017).

\bibitem{wang2018jnd}
Wang, H., Zhang, X., Yang, C., and Kuo, C.-C.~J., ``A {JND}-based video quality
  assessment model and its application,'' {\em arXiv preprint arXiv:1807.00920}
   (2018).

\end{thebibliography}
\bibliographystyle{spiebib} 

\end{document}